# Hydrostatic strain enhancement in laterally confined SiGe nano-stripes

G. M. Vanacore[1†], M. Chaigneau[2], N. Barrett[3], M. Bollani[4], F. Boioli[5], M. Salvalaglio[5], F. Montalenti[5], N. Manini[6], L. Caramella[6], P. Biagioni[1], D. Chrastina[7], G. Isella[7], O. Renault[8], M. Zani[1], R. Sordan[7], G. Onida[6], R. Ossikovski[2], H.-J. Drouhin[9], and A. Tagliaferri[1*]

[1] CNISM and Dipartimento di Fisica, Politecnico di Milano, I-20133 Milano, Italy.
[2] LPICM, Ecole Polytechnique, CNRS, F-91128 Palaiseau, France.
[3] CEA Saclay, CEA DSM IRAMIS SPCSI, F-91191 Gif Sur Yvette, France.
[4] CNR-IFN e L-NESS I-22100 Como, Italy.
[5] Dipartimento di Scienza dei Materiali, Università Milano Bicocca, I-20100 Milano, Italy.
[6] Dipartimento di Fisica, Università degli Studi di Milano, Via Celoria 16, I-20133 Milano, Italy.
[7] CNISM e L-NESS, Dipartimento di Fisica, Politecnico di Milano, I-22100 Como, Italy.
[8] CEA, LETI, MINATEC Campus, 38054 Grenoble Cedex 9, France.
[9] LSI, Ecole Polytechnique, CNRS, F-91128 Palaiseau, France.

[†] Present address: Physical Biology Center for Ultrafast Science and Technology, Arthur Amos Noyes Laboratory of Chemical Physics, California Institute of Technology, Pasadena (CA), United States.
*corresponding author: *alberto.tagliaferri@fisi.polimi.it*

PACS number(s): 81.07.-b; 81.16.Nd; 68.35.Gy; 73.30.+y; 68.37.Uv; 78.30.Am; 68.37.Xy

**Abstract**

Strain-engineering in SiGe nanostructures is fundamental for the design of optoelectronic devices at the nanoscale. Here we explore a new strategy, where SiGe structures are laterally confined by the Si substrate, to obtain high tensile strain avoiding the use of external stressors, and thus improving the scalability. Spectro-microscopy techniques, finite element method simulations and *ab initio* calculations are used to investigate the strain state of laterally confined Ge-rich SiGe nano-stripes. Strain information is obtained by tip enhanced Raman spectroscopy with an unprecedented lateral resolution of ~ 30 nm. The nano-stripes exhibit a large tensile hydrostatic strain component, which is maximum at the center of the top free surface, and becomes very small at the edges. The maximum lattice deformation is larger than the typical values of thermally relaxed Ge/Si(001) layers. This strain enhancement originates from a frustrated relaxation in the out-of-plane direction, resulting from the combination of the lateral confinement induced by the substrate side walls and the plastic relaxation of the misfit strain in the (001) plane at the SiGe/Si interface. The effect of this tensile lattice deformation at the stripe surface is probed by work function mapping, performed with a spatial resolution better than 100 nm using X-ray photoelectron emission microscopy. The nano-stripes exhibit a positive work function shift with respect to a bulk SiGe alloy, quantitatively confirmed by electronic structure calculations of tensile strained configurations. The present results have a potential impact on the design of optoelectronic devices at a nanometer length scale.

**I. INTRODUCTION**

The introduction of SiGe heterostructures into main-stream Si technology has been identified as a possible solution to overcome the physical limitations of Si by opening new degrees of freedom *via* band structure engineering.[1,2] Much work has also been devoted to find the best deposition/fabrication strategy to apply tensile strain to pure Ge structures in order to reduce the energy difference, $\Delta_{gap} = E_{gap}^{dir} - E_{gap}^{indir} = \mathbf{140\ meV}$, between the direct $E_{gap}^{dir}$ and indirect $E_{gap}^{indir}$ band-gaps,[3,4,5] thus favoring population inversion and eventually lasing.[6] A large tensile strain is necessary for enhanced gap shrinkage and, for the same lattice deformation, biaxial strain is more effective than uniaxial strain,[7] possibly due to the larger hydrostatic component which significantly affects the shift of the valence and conduction band states. According to recent **k·p** calculations,[7] Ge grown along the [001] direction acquires a direct band gap (i.e. $\Delta_{gap} = \mathbf{0}$) at around 1.7 % biaxial strain (hydrostatic strain, $\varepsilon_h$ ~ 0.73 %) or around 4.6 % uniaxial strain ($\varepsilon_h$ ~ 0.74 %). The direct gap condition is thus obtained when the biaxial and the uniaxial configurations reach nearly the same hydrostatic strain,[7] suggesting that this is the dominant component for the band gap narrowing.

Different strained configurations have been explored in the literature. A two-dimensional (2D) Ge thin film on Si has a thermally induced tensile biaxial strain of ~ 0.23 % ($\varepsilon_h$ ~ 0.1 %) leading to a reduction of the difference between the direct and indirect band gaps, $\Delta_{gap}$, by ~ 20 meV.[8] A higher tensile deformation has been reached using external stressors. A silicon nitride layer has been used by de Kersauson *et al.*[9] obtaining a tensile biaxial strain of ~ 0.4 % ($\varepsilon_h$ ~ 0.17 %) in 1 μm wide Ge wires, with an optical recombination at around 1690 nm ($\Delta_{gap}$ reduced by 34.7 meV). Recently, Nam *et al.*[10,11] used tungsten as material for the stressor layer to induce a biaxial tensile strain of 0.76 % ($\varepsilon_h$ ~ 0.33 %) and 1.13 % ($\varepsilon_h$ ~ 0.49 %) in 200 μm wide Ge mesas showing light emission at 1710 nm ($\Delta_{gap}$ reduced by ~ 66 meV) and 1750 ($\Delta_{gap}$ reduced by ~ 98 meV), respectively. The works reported at the Refs. 8, 9, 10, and 11 show a clear trend: an enhanced gap shrinkage towards the direct band gap condition can be obtained by increasing the hydrostatic strain component. However, high tensile strain with external stressors has been reached only using thick stressor layers (0.5-1 μm) and for basically large structures (> 1 μm). This reduces considerably the scalability and compromises the application of these methods to the design of optoelectronic devices at a nanometer length scale.

In this paper, we use a strategy to obtain nanoscale structures with a high hydrostatic strain component, avoiding the use of external stressors and thus, in principle, ready to follow the continued downscaling of SiGe heterostructures to increase the performances of integrated circuits. We use epitaxial deposition of Ge on a pre-patterned Si surface with 150 nm wide trenches to create laterally confined Ge-rich SiGe nano-stripes with a microscopic strain state able to maximize the hydrostatic strain component. Nanoscale resolved spectroscopic experiments, finite element method simulations and *ab initio* calculations have been used to map the strain field within the nano-stripes, with an unprecedented lateral resolution of ~ 30 nm. A large tensile hydrostatic strain ($\varepsilon_h$ ~ 0.53 %) is found, as the result of a frustrated relaxation in the out-of-plane direction, that is attributed to the geometrical constraints combined with the plastic relaxation of the misfit strain in the (001) plane. The measured strain is larger than the typical thermal strain in Ge thin films on Si(001) structures ($\varepsilon_h$ ~ 0.1 %) and it is reached without using external stressors. The effect of the lattice deformation at the stripe surface is probed by work function mapping with a spatial resolution better than 100 nm. The fitting of the work function results with electronic structure calculations of strained configurations, provides a quantitative confirmation of the high tensile strain created inside the stripe. The present results have a potential impact on the design of optoelectronic devices at a nanometer length scale for the achievement of band gap narrowing and, eventually, direct gap condition for lasing in SiGe technology.

## II. EXPERIMENTAL DETAILS

### II.A Sample fabrication

SiGe nano-stripes have been fabricated by low energy plasma enhanced chemical vapour deposition (LEPECVD)[12] of Ge on Si substrates patterned by electron-beam lithography (EBL).[13] A n+type (As-doped) Si(001) substrate was patterned with a series of trenches (depth ~ 110 nm, width ~ 150 nm, period ~ 1 μm) aligned along the [110] direction by means of EBL. Then, we epitaxially deposited 15 nm of pure Ge by LEPECVD with a growth rate of 1.5 nm/s at a substrate temperature of 650 °C. Under these growth conditions, the migration length of Ge adatoms is much longer than the separation between the nano-stripes.[14] This favors the gathering of Ge from the surrounding surface area into the trenches, which behave as material traps and represent preferential nucleation sites since a total elastic energy minimum is reached at their base.[15] The whole process leads to the formation of laterally confined nano-stripes. Any SiGe epitaxial layer formed in between the structures was completely etched away by a gently mechanical polishing performed after the Ge growth. The nano-stripes exhibit a lateral width of ~ 150 nm and a thickness of ~ 110 ± 5 nm as determined by several cross-sectional SEM images after focused ion beam (FIB) processing (see Fig. 1). It is worth noting that the high growth rate (1.5 nm/s), the moderate substrate temperature (650 °C), and the very short deposition time (10 s) have been used in order to strongly reduce the Si incorporation from the substrate, leading to the formation of Ge-rich nanostructures.[16]

### II.B Tip Enhanced Raman Scattering (TERS)

The TERS setup consists of a Horiba Jobin Yvon Labram HR800 Raman Spectrometer optically coupled in an oblique backscattering geometry (65° with respect to the sample normal) to a Park Systems XE-100 Scanning Tunnelling Microscope through a long-working-distance objective (50×, numerical aperture of 0.45). The opto-mechanical coupling is motorized along the *x*, *y* and *z* axes and allows for an accurate automated alignment of the exciting light spot with respect to the tip apex in near-field scattering (TERS) experiments. The excitation wavelength of 633 nm is provided by the built-in HeNe laser of the spectrometer. The polarization state of the incident radiation was set at *p*-polarization (electric field parallel to the scattering plane) by using a half-wave plate inserted in motorized rotating mount. STM tips were prepared by electrochemical etching from 0.25 mm diameter Au wire (Goodfellow) in a concentrated HCl/ethanol 1:1 mixture.[17,18] Tips with final apex radius lower than 30 nm can be reproducibly fabricated using this technique.[19] Tunneling experiments took place in air using a 1V tip-positive sample bias voltage and a 0.1 nA tunneling current. Before every measurement sequence, native silicon and germanium oxides on the sample surface have been removed by means of a diluted HF solution (10 % for 30 s at RT).

**II.C X-Ray PhotoElectron Emission Microscopy (X-PEEM)**

The XPEEM experiments were carried out at the TEMPO beamline of SOLEIL Synchrotron using a fully energy-filtered PEEM instrument (NanoESCA, Omicron Nanotechnology).[20] This is composed of an electrostatic PEEM column together with an energy filter consisting of two hemispherical electron energy analyzers coupled by a transfer lens. Soft X-rays with 90 eV and 160 eV photon energy have been used for core-level and work function mapping. The sample was mounted such that the normal to the (001) surface was in the horizontal scattering plane containing the incoming wavevector. The light was incident at an angle of 23° with respect to the (001) plane, and horizontal linear polarization of the incident light was chosen in order to have a preferential sensitivity along the [001] out-of-plane direction. The NanoESCA spectro-microscope was operated with a contrast aperture of 70 μm, an extractor voltage of 15 kV, a pass energy of 100 eV, and an entrance analyzer aperture of 1 mm, giving a spectrometer resolution of 0.4 eV. A field of view (FoV) of 15 μm was used. All images were corrected for the inherent non-isochromaticity.[21] Dark and flat field corrections for camera noise and detector inhomogeneities were also applied. A four stage preparation protocol for the cleaning of the sample surface was used: (*i*) chemical etching of the native silicon and germanium oxide by diluted HF (10 % for 30 s at RT); (*ii*) UV-ozone treatment by irradiation with $D_2$ lamp under $O_2$ flux (15 ÷ 20 min) for carbon removal;[22,23] (*iii*) removal of silicon oxide layer (covering the surface after the UV treatment) by *in-situ* mild $Ar^+$ sputtering (beam voltage ~ 500 V ÷ 1000 V, beam current ~ 1 μA), and (*iv*) thermal relaxation of residual sputtering damage by *in situ* annealing below the diffusion threshold temperature (~ 400 °C). Fig. 2(a) reports the monitoring of the surface contaminations during the different steps of the cleaning procedure on a Si test sample using Auger electron spectroscopy, while in Fig. 2(b) is shown the photoemission spectrum of the Ge 3d core level measured on the nano-stripes after the cleaning procedure. This indicates the absence of germanium-oxide contamination since no chemical shifted structures appears at the low kinetic energy (defined by $E$-$E_F$, the electron energy measured with respect to the sample Fermi level, $E_F$) side of the spectrum.

## III. RESULTS AND DISCUSSION

### III.A Nanoscale strain mapping

Strain information is obtained by Tip Enhanced Raman Spectroscopy (TERS), allowing for Raman spectroscopic imaging with high spatial resolution at the surface of the sample.[24,25] The approach presents unique advantages compared to other techniques, namely nanostructural investigation by transmission electron microscopy (TEM)[26,27] and nanobeam X-ray diffraction (NXRD)[13]. Indeed, although diffraction by TEM has been shown to provide in specific cases detailed strain information down to the nanometer scale,[26] it is also a destructive technique that would require invasive thinning of the samples down to a length scale comparable to the size of the structure of interest. This procedure would induce a significant elastic and possibly plastic relaxation that would make the reconstruction of the initial strain state difficult and uncertain.[27] Conversely, although NXRD shares with TERS the non-destructive character, it is limited to a lateral resolution of few hundred nm and is a bulk sensitive technique, unable to provide the surface information that is relevant for the optoelectronic application mentioned above.

TERS exploits the local amplification of the electromagnetic field at the apex of a sharp gold tip, stabilized by feedback control of the tunneling current between the tip and the sample.[28] This converts the incoming far-field radiation from a focalized laser beam into an enhanced near-field, spatially confined in a region whose extent is roughly of the order of the tip apex radius (20-30 nm, with an improvement of the incident intensity of more than an order of magnitude with respect to far field Raman investigations).[26] The incident radiation is $p$-polarized (electric field parallel to the scattering plane) with the [110] crystallographic axis laying within the scattering plane, as schematically shown in Fig. 3(a), which maximizes both the TERS enhancement[29] and the contrast between near-field and far-field[30]. A region containing a single nano-stripe is selected by Scanning Tunneling Microscopy (STM) imaging of the surface (see Fig. 3(b)), and the TERS measurement is performed by scanning the tip over the chosen stripe.

Fig. 3(c) shows the baseline-corrected TERS spectra measured with the tip in tunneling position on the nano-stripe, and on the Si substrate. The most intense peaks related to the far-field contribution from the Si bulk (the Si-Si 1$^{st}$ order mode at 520.7 cm$^{-1}$ and the Si-Si 2TA overtone at 300 cm$^{-1}$) are almost unchanged for the two tip positions. Over the nano-stripe additional features appear in the spectrum: a doublet structure at 553-575 cm$^{-1}$, and a well-resolved peak at 380 cm$^{-1}$. These findings can be understood considering that the far-field radiation probes the same large scattering volume in the bulk Si even when the tip is on the small Ge stripe, whereas the locally enhanced near-field component probes only the nano-stripe (see the small light green hemispheres in the schematics in the inset of Fig. 3(c)). The doublet structure at 553-575 cm$^{-1}$ is attributed to the 2TO overtones at W and L, respectively, of the Ge-Ge Raman mode,[31,32] while the 380 cm$^{-1}$ peak is assigned to the 1$^{st}$ order component of the Si-Ge Raman mode[31,33] of the nano-stripe due to long-range order lattice vibrations. Concerning this attribution for the 380 cm$^{-1}$ peak, notice that the 2LA Ge-Ge overtone[32] at 382 cm$^{-1}$ is negligible. In fact both theory[34] and experiments[32,35] report that the Ge-Ge 2TO peaks are six times more intense than the Ge-Ge 2LA component. In our spectra the intensity of the peak at ~ 380 cm$^{-1}$ is always greater by one order of magnitude than the 2TO Ge-Ge overtone. Moreover, as measured by photoelectron emission microscopy (see below), the stripes are rich in Ge and for high Ge content alloys the Si-Ge peak has been theoretically predicted and experimentally found in the range 380-390 cm$^{-1}$.[16,36]

Figs. 3(d) and 3(e) show representative intensity profiles of the Ge-Ge 2TO and of the 1$^{st}$ order Si-Ge peaks as a function of the position across a single nano-stripe with lateral resolution ~ 30 nm. The enhancement due to the near-field contribution, and the main trend of the profiles, are reproducible within the experimental uncertainty and are consistent with the stripe width of 150 nm. The Si-Ge Raman peak monitored across the stripe moves from higher to lower frequencies as the tip is moved from the side toward the center of the stripe (see Fig. 3(f)).

In heteroepitaxial SiGe structures the frequency of the Raman peaks is strongly dependent on strain and Ge content.[31,33] The composition of the nano-stripes is obtained by X-ray photoelectron emission microscopy (XPEEM). The Ge concentration is measured by acquiring energy filtered photoelectrons image series of the Ge 3d (binding energy, $E_B$ ~ 29 eV) and Si 2p ($E_B$ ~ 99 eV) core levels, at the same kinetic energy of the emitted electrons (~ 61 eV for excitation with 90 eV and 160 eV photon energy, respectively). These XPEEM measurements are recorded as three-dimensional (3D) data sets of the photoemission intensity $I(E_K, x, y)$ as a function of the kinetic energy, $E_K$, and of the position, $x$ and $y$, within the field of view. The main panels in Figs. 4(a) and 4(b) represent the background subtracted core level images, showing the photoemitted intensity at the peak energy of the Ge 3d and the Si 2p core levels, respectively. The insets show the spectra extracted from the 3D data set by averaging the photoemission signal at a given energy over a single nano-stripe. The spatial map of the Ge concentration in Fig. 4(c) is then obtained by fitting the local intensities of the Si 2p and Ge 3d core-levels using a standard quantification model (see Appendix A for further details). The nano-stripes are Ge-rich and exhibit an almost square concentration profile with an average Ge concentration $x_{Ge} \approx 91 \pm 3\%$. This evaluation is strictly valid for the region at the center of the stripe with an extension of the order of the XPEEM spatial resolution (~ 100 nm), whereas variations of the concentration larger than the experimental uncertainty cannot be excluded towards the stripe walls boundary, as discussed later.

The strain state of the nano-stripes can be now obtained by fitting the measured frequency[31] of the Si-Ge Raman peak at 380 cm$^{-1}$ with the expected strain-induced frequency shift for the employed TERS configuration, evaluated using the alloy composition obtained with XPEEM. The 1$^{st}$ order Si-Ge peak has been used in order to rule out possible non-linear effects present in the multi-phonon processes, that can conversely play a role in the 2$^{nd}$ order Ge-Ge spectral feature.

In diamond-lattice crystals the first-order $q = 0$ optical phonon is triply degenerate with two transverse (TO1 and TO2) modes and one longitudinal (LO) mode. Symmetry breaking by the lattice distortion lifts the degeneracy resulting in frequency splitting and modification of the Raman polarizability tensors of the three modes[37] (see Appendix C for a detailed description). In confocal backscattered far-field Raman spectroscopy only the longitudinal mode is excited.[38] In the present near-field TERS experiment, the Raman polarizability tensors are also modified by the presence of the tip leading to changes in the selection rules. Following the model by Ossikovski et al.[39] we consider a "tip-amplification tensor", accounting for the interaction between the tip and the electromagnetic field, and determine the scattered intensities for the three phonon modes as a function of the angle, $\vartheta$, between the light direction and the tip axis:

$$I_{TO1} = I_{TO2} \propto (ab)^2 (2\sin^2 \vartheta + \cos^2 \vartheta) \qquad (1.a)$$
$$I_{LO} \propto b^4 \cos^2 \vartheta \qquad (1.b)$$

where $a$ and $b$ are phenomenological tip-amplification factors (with $a > b$) related to the longitudinal and transverse tip polarizability, respectively. Considering $\vartheta = 65°$ and a typical experimental value of $a/b \approx 5.5$ for the employed tip,[30] the ratio $I_{LO}/I_{TO}$ is $\approx 10^{-3}$, demonstrating that the TERS signal in the present backscattered oblique configuration is dominated by the two TO modes, which are not normally accessible in a standard confocal Raman experiment where the LO mode dominates. This analysis shows that a TERS experiment can give access to new information otherwise prohibited by the selection rules. The same qualitative and quantitative results are obtained by calculating Raman selection rules in the framework of the electromagnetic theory of near-field Raman enhancement (see Appendix C). Since the transverse modes TO1 and TO2 are excited with the same probability and their splitting cannot be resolved, the strain-induced frequency shift of the first order $q = 0$ optical phonon peaks in the measured TERS spectra, $\Delta\omega$, can be obtained by the average of the TO1 and TO2 mode frequencies (see Appendix C). This leads to the following expression:

$$\Delta\omega \approx \frac{3}{4}\omega_0(x_{Ge})[K_{11}(x_{Ge}) + K_{12}(x_{Ge})]\varepsilon_h \qquad (2)$$

which relates the Raman frequency shift $\Delta\omega$ and the hydrostatic strain component $\varepsilon_h = \frac{1}{3}\text{Tr}\{\boldsymbol{\varepsilon}\}$ (where $\boldsymbol{\varepsilon}$ is the strain tensor and $\text{Tr}\{...\}$ is the trace operator) of the investigated structures. In the Eq. (2), $\omega_0(...)$ is the composition-dependent unstrained mode frequency, $K_{11}(...)$ and $K_{12}(...)$ are the composition-dependent phonon deformation potentials (PDP), and $x_{Ge}$ is the Ge concentration in the stripe.

Several authors[40,41] reported that for $x_{Ge}$ larger than 50%, the Si-Ge peak shifts to lower frequency values as the strain changes from compressive (negative) to tensile (positive). Thus the red shift of the Si-Ge peak, measured while moving from the edge to the center of the nano-stripe (see Fig. 3(f)), is consistent with an increasing tensile distortion. The hydrostatic strain, $\varepsilon_h$, within the stripe can thus be obtained from the Eq. 2 using the Si-Ge peak frequency trend determined by the TERS analysis. The parameters $\omega_0^{Si-Ge}$, $K_{11}^{Si-Ge}$ and $K_{12}^{Si-Ge}$ for the Si-Ge peak at 380 cm$^{-1}$ are determined from the Ge concentration of the stripe, as measured by XPEEM, using their composition dependence reported in Ref. 41, where bulk-like values for the PDPs are calculated. As it will be discussed below, the use of bulk-PDPs leads to strain values which will be satisfactorily reproduced by the simulations of the strain field inside the stripe. Within the range of values that has been reported in literature for the set of the $\omega_0$, $K_{11}$ and $K_{12}$ parameters, the relations provided by Ref. 41 result in the lowest and most conservative strain values. The two strain profiles in Fig. 5(a) represent the results from two line scan at different positions along the stripe axis. The profiles have similar shape and intensity. The local differences around 30 nm are attributed to morphological changes in the section of the stripe at the scan positions.

**III.B Size-dependent frustrated lattice deformation**

The origin of the measured $\varepsilon_h$ profile across a single nano-stripe is now discussed. The application to the TERS data of the strain-induced frequency shift described by the Eq. (2) and based on the Ossikovski's model, gives (see Fig. 5(a)) a tensile hydrostatic strain, $\varepsilon_h$, that is maximum (~ $5.3 \times 10^{-3}$) at the center of the stripe and almost vanishes at the edges. Far from the edges moving toward the stripe center, strain values are noticeably higher than the typical hydrostatic thermal strain in a 2D film ($\varepsilon_h$ ~ 0.1%). A better understanding of the origin of such tensile deformation can be obtained by applying elasticity theory using finite element methods (FEM). An idealized geometry (see Fig. 5(b)) is considered for the simulations, where the stripe section is a perfect rectangle of width 150 nm (along the *x* direction) and height 110 nm (z-axis), corresponding to the experimental geometry. The exact shape of the stripe does not affect significantly the hydrostatic strain $\varepsilon_h$ at its surface, provided that the same aspect ratio and section is preserved (as measured in Fig. 1(b)). The robustness of this approach has been carefully verified by comparing the $\varepsilon_h$ values calculated by FEM over a representative set of different stripe sections, characterized by rounded, smooth or rough boundaries (see Appendix E). The size of the stripe in the third direction (y-axis) is considered infinite, allowing us to perform simpler 2D calculations.

As the average Ge content is measured to be extremely high ($x_{Ge}$ ~ 91%), the height of the stripe is almost two orders of magnitude larger than the critical thickness for misfit-dislocation insertion in a film with the corresponding Ge content.[42] The simplest meaningful approach has been adopted to model the effect of the plastic relaxation. In the *y* direction, the strain is expected to relax in a film-like mode due to the infinite extension of the stripe. Thus in the simulations the initial $\varepsilon_{yy}$ value was set equal to the thermal strain obtained in a film $\varepsilon_{yy} = +0.0023\, x_{Ge}$,[8] i.e. a full plastic relaxation of the epitaxial misfit has been assumed. The actual thermal-strain value influences the results only slightly, the main effect comes from the $\varepsilon_{zz}$ strain relaxation arising from the geometrical constrictions along *x*, as described in the following.

In order to tackle the strain relaxation along the *x* direction, expected to significantly deviate from the film-like behavior,[43] an ordered array of straight edge dislocations has been explicitly considered to extend in the *y* direction (thus, relaxing the strain along the *x* direction), and placed at the bottom interface between the stripe and the Si substrate (see the dislocation-induced strain field in the inset of Fig. 5(b)). The technique used to treat dislocations within FEM is described in Ref. 44. After verifying that the system energy is minimized by a number $N_d$ of such dislocations (the energy curve as a function of $N_d$, not shown, presents a flat minimum, with values within the error for $N_d$ equal to 14 and 15 in the present case), we computed the $\varepsilon_h$ strain component, displayed in Fig. 5(b). As the misfit in the *x* direction is partially removed plastically (the top free surface allowing for some extra elastic relaxation), the main stress to which the stripe reacts is along the *z* direction, due to the vertical walls boundary with the Si substrate, that tends to reduce the stripe lattice to the bulk Si one. The boundary region closer to the free surface is deformed more easily. As a result, the $\varepsilon_h$ strain map is strongly non-homogenous, top portions becoming tensile-strained in order to allow for a better relaxation of the whole system.

In order to compare FEM simulations and experiment, the expected TERS results have been simulated for the ideal stripe with a strain state as obtained from the theoretical calculations. The TERS simulation has taken into account the TERS probing volume by performing a weighted average of the $\varepsilon_h$ values over it. The probing volume has been quantified from the lateral resolution, experimentally determined (see Fig. 3(d-e)), and from the near-field penetration inside the stripe, as simulated by Full-field 3D finite-difference time-domain (FDTD) calculations.[45]. The FDTD simulations have been made using the following parameters: the dielectric constants for SiGe and Au are derived from the literature;[46,47] and a Gaussian beam was used to illuminate the tip, having the same wavelength and focusing parameters as in the experimental setup. The results of the FDTD simulation are presented in Fig. 6(a), that shows the intensity of total electric field in the *xz* plane after interaction with a gold tip having a radius of 20 nm and separated of 0.7 nm from the *z* = 0 plane (the (001) plane), while the Fig. 6(b) shows the Raman signal (proportional to the fourth power of the electric field) in the $Si_{0.1}Ge_{0.9}$ alloy together with the transversal and longitudinal profiles. The attenuation of the TERS signal shows a penetration depth of about 3.1 nm.

The strain values obtained by the FEM simulations have been then weighted along the z-axis with the attenuation curve of the near-field inside the stripe, as calculated by FDTD simulations, and convoluted along *x* with a Gaussian function of 30 nm full-width at half-maximum, leading to the continuous curve displayed in Fig. 5(a). The comparison with experiments is fairly satisfactory, in the sense that high values of $\varepsilon_h$ are found (some 3 times higher than the maximum expectation for a flat film). However, there is a tendency of FEM simulations to underestimate the strain at the center of the stripe. The difference is most probably due to the simple approach used for treating plastic relaxation, while the oversimplified rectangular stripe shape assumed in the calculations to approximate the real, smoother profile mainly affects the strain at the stripe edges rather than at the center. The present simulation has thus to be considered only a first step in the comprehension of the complex network of defects that is determining the final strain state, the full process of plastic relaxation in such a structure still needs to be clarified, possibly involving also dislocations running along the stripe walls. Nevertheless, the present FEM results show that $\varepsilon_h$ values (locally) larger than the film case are possible in nanoscale structure without using external stressors, and that the tensile strain has a different origin than the 2D geometry of a film and larger microscale structures (thickness dependent plastic, elastic and thermal relaxation processes), where it is strictly related to the local lateral constraints.

**III.C Strain-induced work function changes**

The tensile strain field inside the shallow volume of the stripe is responsible for a strong modification of the surface electronic band structure, in particular for the shift and splitting of the conduction and valence bands. Recent experiments[48,49] and *ab initio* calculations[50] pointed out that the modifications of the valence band induced by the strain are correlated with changes in the work function. These changes result from the modification of the surface electrostatic dipole,[50,51] and from the shift of the Fermi level due to the band structure warping as the lattice is deformed.[52] No evidences of quantum confinement effects on the electronic structure have been reported at the length scales investigated.[53,54] Here we test these concepts by spatially mapping the work function of the nano-stripes and provide a further quantitative confirmation of the lattice deformation created at their surface.

The work function is measured by acquiring XPEEM image series at the photoemission threshold. The spectra as a function of the final state energy referred to the Fermi level, $E$–$E_F$, are characterized by a sharp threshold depending on the local work function of the emitting region.[55] Figs. 7(a) and 7(b) represent the laterally resolved photoemitted intensity using secondary electrons at a kinetic energy of 4.6 eV and 4.9 eV, respectively, where contrast inversion between the nano-stripes and the surrounding Si bulk reflect different work function values. The work-function map in Fig. 7(c) is obtained from the experimental pixel-by-pixel threshold spectra least-squares fitted to the full secondary electron distribution model described by Henke *et al.*[56] (see Appendix B). After correction for the Schottky effect taking into account the energy change of electrons extracted by an immersion lens (0.11 eV for an extraction voltage of 15 kV),[57] we obtain $\Phi_{substrate} = 4.66 \pm 0.02$ eV at the Si substrate surface and $\Phi_{stripe} = 4.85 \pm 0.03$ eV inside the SiGe stripe surface. The value for the substrate is similar to that reported for n-type doped Si(001).[58] The work function value of the SiGe stripe is larger than that of a bulk $Si_{0.1}Ge_{0.9}$ alloy ($4.768 \pm 0.015$ eV)[59] by ~ 80 meV.

The work function increase is consistent with a tensile deformation at the surface of the stripe. In fact, under a tensile strain the potential deformation theory[60] predicts a significant decrease in the energy of the conduction band minima for both Γ and L valleys, and a slight increase of the valence band edge. This leads to a narrowing of the direct (in Γ) and indirect (in L) band gaps, and to a lowering of the Fermi level yielding a larger work function. A further contribution is the strain-induced modification of the surface electrostatic dipole, which is predicted to decrease the work function in case of tensile strain (i.e. to have an opposite sign with respect to the band structure-induced variation): under tensile deformation a lower electronic charge density is distributed outside the surface reducing the dipole strength and thus lowering the work function. However, our experimental findings of an increase of the nano-stripes work function compared to an unstrained bulk alloy, are consistent with and supported by recent *ab initio* calculations,[50] showing that the magnitude of the Fermi level shift is greater than that of the surface dipole strength.

To obtain a quantitative estimation of the tensile deformation present at the surface of the nano-stripes from the experimental work-function shift, density functional theory (DFT) calculations under the local-density approximation (LDA) have been performed, computing the band structure and the surface electrostatic potential of a strained Ge(001) slab. The tensile deformation along the main symmetry directions has been applied by increasing the cell sides. Details of the calculations are reported in the Appendix D. For a tensile strain, a downward shift of both the Fermi level and the conduction band is found, together with a reduction of the surface electrostatic dipole, partly cancelling the effect of the conduction band lowering. Figs. 8(a)-(d) shows the calculated band structure of the Ge(001)b(2×1) surface for the unstrained case (panel (a)), and for a tensile strain of 1 % separately applied along the *x*-axis (panel (b)), *y*-axis (panel (c)) and the *z*-axis (panel (d)). The blue region in Fig. 8(e) represent the calculated values of the work function shift, $\Delta\Phi$, as a function of the hydrostatic strain, $\varepsilon_h$, considering the possible combinations for the non-zero components of the strain tensor ($\varepsilon_{xx}$, $\varepsilon_{yy}$, and $\varepsilon_{zz}$) consistent with the values obtained by the FEM simulations: $\varepsilon_{yy} = \mathbf{0.0023}\, x_{Ge}$ (film-like thermal strain due to full plastic relaxation along *y*), $0 < \varepsilon_{xx} < 0.02$ (tensile strain due to the plastic relaxation along *x* and the compressive load along *z*), $-0.006 < \varepsilon_{zz} < 0$ (compressive strain introduced by the vertical walls along z, attempting to enforce the Si lattice parameter). By comparing the estimated work-function shift of ~ 80 meV with the calculated values, the hydrostatic strain at the stripe surface is estimated to range from ~ 0.4 % to 0.65 %, in substantial agreement with the value obtained from the TERS analysis (~ 0.53 %).

## IV. CONCLUSIONS AND OUTLOOK

In this paper we presented the nanoscale mapping of composition and strain at the surface of SiGe nano-stripes laterally confined in 150 nm wide Si trenches. Such stripes exhibit a high Ge concentration (~ 91 %), a positive shift of the work function (~ 80 meV), and a tensile strain which is maximum at the center ($\varepsilon_h \sim \mathbf{0.53}$ %) and becomes very small at the edges. This strain behaviour is understood as the result of a frustrated relaxation in the out-of-plane direction due to the constrained geometry, combined with the plastic relaxation of the misfit strain in the (001) plane, leading to a tensile deformation at the top of the stripe. The positive work-function shift is attributed to the warping of the surface electronic structure as induced by a tensile lattice deformation, providing a further quantitative confirmation of the strain state of the nano-stripes.

Our results represent the experimental demonstration that a high hydrostatic tensile strain can be achieved in nanoscale structures without using external stressor layers. A future experimental investigation of their optical properties is envisaged in order to probe their actual applicability in SiGe laser technology. The task is particularly demanding because state-of-the-art far-field optical techniques, such as absorption, ellipsometry, or photoluminescence provide volume averaged information over the whole nanostructure volume, with very low sensitivity to the small strain volume at surface.

The approach used in this work shows promising possibilities for the achievement of a band gap narrowing and, eventually, a direct gap condition for lasing, provided that a reduction of the Si incorporation is obtained and strategies for the quenching of defects formation are explored. Moreover, an enhancement of the strain is reasonably achievable through the exploration of different trench geometries. This makes the selective growth of Ge on a Si substrate patterned with seeding structures a technological pathway to Silicon compatible and scalable process for the design of optoelectronic devices at a nanometer length scale.

## ACKNOWLEDGEMENTS

Special thanks are due to J. Leroy, B. Delomez, M. Lavayssierre, J. Rault, C. Mathieu, F. Sirotti, M. Silly for technical assistance during XPEEM experiments, E. Gualtieri for the FIB measurements, E. Bonera for valuable discussions and S. Collin for preliminary optical micro-reflectivity measurements at CNRS/LPN, Marcoussis, France. The work was supported by the French National Research Agency (ANR) through the "Recherche Technologique de Base (RTB)" program. We acknowledge SOLEIL for provision of synchrotron radiation, and the ETSF for theoretical support through the user project n. 458.

**APPENDIX A: Ge CONCENTRATION MAPPING**

The Ge concentration within the nano-stripes has been measured by acquiring energy filtered image series around the Ge 3d (binding energy, $E_B \sim 29$ eV) and Si 2p ($E_B \sim 99$ eV) core levels using soft x-ray excitation at 90 eV and 160 eV, respectively. The local Ge concentration can be obtained from the total intensities using the usual relation:[61]

$$I_i = (J_0 \sigma \lambda T) x_i$$

where the subscript $i$ stands for Si or Ge, $J_0$ is the photon flux, $\sigma$ the photoemission cross-section, $\lambda$ the inelastic mean free path ($\sim 0.5$ nm in high Ge content SiGe alloy for photoelectrons at 60 eV),[62] $x_i$ the elemental concentration and $T$ the analyzer transmission. The choice of the photon energies means that the kinetic energy (and thus $\lambda$ and $T$) are the same for both Si 2p and Ge 3d core levels. The Ge concentration can be thus readily obtained by:

$$x_{Ge} = \frac{\frac{I_{Ge}}{J_0(h\nu)\sigma_{Ge}(h\nu)}}{\frac{I_{Ge}}{J_0(h\nu)\sigma_{Ge}(h\nu)} + \frac{I_{Si}}{J_0(h\nu)\sigma_{Si}(h\nu')}}$$

where $h\nu = 90$ eV and $h\nu' = 160$ eV. Using Yeh and Lindau's cross-sections[63] and the known response of the x-ray monochromator at the TEMPO beamline, the spatial map of the Ge concentration, shown in Fig. 4(c), is obtained. This is actually the convolution between the *real* concentration map and a Gaussian function with a full width half maximum (FWHM) determined by the lateral resolution of the microscope. De-convolution of the experimentally measured concentration profile across a single 150 nm wide stripe, using the combination of a Van Cittert method[64] and a Lanczos filter[65], allowed to estimate a spatial resolution for core level mapping of $96.7 \pm 3.5$ nm. The nano-stripes are rich in Ge and exhibit an almost square concentration profile with an average Ge concentration of $0.91 \pm 0.026$.

**APPENDIX B: WORK FUNCTION MAPPING**

The work function, $\Phi = E_0 - E_F$, where $E_0$ is the vacuum level and $E_F$ is the Fermi level of the sample surface, has been measured by acquiring XPEEM image series at the photoemission threshold. Fig. 9 shows the threshold spectra extracted from the Si bulk and from a single nano-stripe. The energy scale on the abscissa axis is represented by the final state energy, $E$, referred to the sample Fermi level, $E_F$.

If $E_K$ denotes the kinetic energy of the photoelectrons measured at the entrance of the imaging analyzer, then we can write that:

$$E - E_F = E_K + \Phi_A,$$

where $\Phi_A$ is the work function of the analyzer. An electron having a binding energy $E_B$ with respect to $E_F$, excited with photons of energy $h\nu$, will have a measured kinetic energy $E_K$ given by:

$$E_K = h\nu - E_B - \Phi_A.$$

At the photoemission threshold we have that $h\nu - E_B = \Phi$, where $\Phi$ is the sample work function. Thus, the threshold measured kinetic energy $E_K^0$ is:

$$E_K^0 = \Phi - \Phi_A,$$

and the correspondent final state energy referred to the Fermi level is:

$$(E - E_F)^0 = \Phi.$$

The secondary electron energy distributions of Fig. 9 are thus characterized by a sharp threshold corresponding to the local work function $\Phi$ of the emitting region under consideration. The local work function map of the Si substrate and of the SiGe nano-stripes (shown in Fig. 7(c)) has been then obtained from the best least-square-fitting of the experimental pixel-by-pixel threshold spectra measured over the field of view to the full secondary electron distribution, $S(E–E_F)$, described by Henke *et al.*:[56]

$$S(E - E_F) = \frac{A(E - E_F - \Phi)}{(E - E_F - \Phi + B)^4}$$

where $A$ is a scaling factor and $B$ is a fitting parameter. The accuracy of $\Phi$ using this procedure has been estimated to be $\pm$ 20 meV.[66]

The de-convolution of the experimentally measured work function profile across a single 150 nm wide stripe from a Gaussian allowed to estimate the spatial resolution to be around 88 nm. The work function value measured for the SiGe stripes is $\Phi_{stripe}$ = 4.85 $\pm$ 0.03 eV, larger than the work function of a bulk $Si_{0.1}Ge_{0.9}$ alloy (4.768 $\pm$ 0.015 eV) obtained by the weighted sum of the intrinsic Si(001) (4.75 eV)[58] and Ge(001) (4.77 eV)[59].

We exclude work function changes due to oxidation of germanium[67] on the SiGe nano-stripes since no chemical shifted structures appears at the low kinetic energy side in the Ge 3d spectrum (see Fig. 2).

**APPENDIX C: STRAIN DETERMINATION FROM TERS**

**C.1 Strain-induced modification of Raman polarizability**

In SiGe structures heteroepitaxially grown on Si substrates the presence of strain due to the lattice deformation between the SiGe alloy and the Si bulk induces a shift of the Raman peaks.[40] In diamond lattice crystals as Si and Ge the $q = 0$ Raman active optical phonon is triply degenerate and is composed by two transversal modes (TO1 and TO2) and one longitudinal mode (LO). The breaking of the cubic symmetry induced by the lattice distortion is able to lift the degeneracy resulting in a frequency splitting and in a modification of the Raman polarizability tensors, $\boldsymbol{R_i}$, of the three modes[37] (where $i = 1$ (TO1), 2 (TO2), and 3 (LO)). $\boldsymbol{R_i}$ tensor have the following matrix representation in the crystal reference framework identified by the [100] ($x'$-axis), [010] ($y'$-axis) and [001] ($z$-axis) directions:

$$\boldsymbol{R_1} = \begin{bmatrix} 0 & 0 & 0 \\ 0 & 0 & 1 \\ 0 & 1 & 0 \end{bmatrix} \quad \boldsymbol{R_2} = \begin{bmatrix} 0 & 0 & 1 \\ 0 & 0 & 0 \\ 1 & 0 & 0 \end{bmatrix} \quad \boldsymbol{R_3} = \begin{bmatrix} 0 & 1 & 0 \\ 1 & 0 & 0 \\ 0 & 0 & 0 \end{bmatrix}.$$

The detailed procedure to get the split mode frequencies and the strain-modified Raman polarizability tensors can be found in a number of references (see Ref 37 for a review). Here we give only a brief overview of it, sufficient for the purpose of the present work.

Due to the stripe morphology and geometry we consider a diagonal representation of the stress tensor experienced by the SiGe material of the stripe within the stripe reference framework identified by the [110] ($x$-axis), [1-10] ($y$-axis) and [001] ($z$-axis) crystallographic directions:

$$\boldsymbol{\varepsilon} = \begin{bmatrix} \varepsilon_{xx} & 0 & 0 \\ 0 & \varepsilon_{yy} & 0 \\ 0 & 0 & \varepsilon_{zz} \end{bmatrix}.$$

The frequency splitting for the $i$th mode, $\omega_i$, is obtained by the eigenvalues $\lambda_i$ of the symmetric matrix $\boldsymbol{K}$,[68] which couples the phonon deformation potentials with the strain components expressed within the crystal reference framework ($x'$:[100], $y'$:[010], $z'$:[001]):

$$K = \begin{bmatrix} p\varepsilon_{x'x'} + q(\varepsilon_{y'y'} + \varepsilon_{z'z'}) & 2r\varepsilon_{x'y'} & 2r\varepsilon_{x'z'} \\ 2r\varepsilon_{x'y'} & p\varepsilon_{y'y'} + q(\varepsilon_{x'x'} + \varepsilon_{z'z'}) & 2r\varepsilon_{y'z'} \\ 2r\varepsilon_{x'z'} & 2r\varepsilon_{y'z'} & p\varepsilon_{z'z'} + q(\varepsilon_{y'y'} + \varepsilon_{x'x'}) \end{bmatrix}$$

where $p$, $q$ and $r$ are the so-called phonon deformation potentials. In order to determine the strain components $\varepsilon_{x'x'}$, $\varepsilon_{y'y'}$, $\varepsilon_{z'z'}$, $\varepsilon_{x'y'}$, $\varepsilon_{x'z'}$, and $\varepsilon_{y'z'}$, the stress tensor $\boldsymbol{\varepsilon}$ expressed in the stripe reference framework ($x$:[110], $y$:[1-10], $z$:[001]) needs to be consequently transformed according to:

$$\boldsymbol{\varepsilon}' = T(\delta)^T \boldsymbol{\varepsilon} T(\delta)$$

where

$$T(\delta) = \begin{bmatrix} \cos\delta & \sin\delta & 0 \\ -\sin\delta & \cos\delta & 0 \\ 0 & 0 & 1 \end{bmatrix}$$

is the corresponding rotation matrix and the rotation angle, $\delta$, is 45° in this case. The rotated strain tensor, $\boldsymbol{\varepsilon}'$, is thus readily obtained:

$$\boldsymbol{\varepsilon}' = \frac{1}{2}\begin{bmatrix} \varepsilon_{xx} + \varepsilon_{yy} & \varepsilon_{xx} - \varepsilon_{yy} & 0 \\ \varepsilon_{xx} - \varepsilon_{yy} & \varepsilon_{xx} + \varepsilon_{yy} & 0 \\ 0 & 0 & 2\varepsilon_{zz} \end{bmatrix},$$

and thus:

$$\varepsilon_{x'x'} = \varepsilon_{y'y'} = \left(\frac{\varepsilon_{xx} + \varepsilon_{yy}}{2}\right)$$

$$\varepsilon_{z'z'} = \varepsilon_{zz}$$

$$\varepsilon_{x'y'} = \left(\frac{\varepsilon_{xx} - \varepsilon_{yy}}{2}\right)$$

$$\varepsilon_{y'z'} = \varepsilon_{x'z'} = 0.$$

The eigenvalues $\lambda_i$ (where $i = 1$ (TO1), 2 (TO2), and 3 (LO)) of the matrix $K$ can be thus directly correlated to the components of the strain components $\varepsilon_{xx}$, $\varepsilon_{yy}$, and $\varepsilon_{zz}$:

$$\lambda_1 = p\varepsilon_{x'x'} + q\varepsilon_{y'y'} + q\varepsilon_{z'z'} + r\varepsilon_{x'y'} = \left(\frac{\varepsilon_{xx} + \varepsilon_{yy}}{2}\right)(p+q) + q\varepsilon_{zz} + r\left(\frac{\varepsilon_{xx} - \varepsilon_{yy}}{2}\right)$$

$$\lambda_2 = p\varepsilon_{x'x'} + q\varepsilon_{y'y'} + q\varepsilon_{z'z'} - r\varepsilon_{x'y'} = \left(\frac{\varepsilon_{xx} + \varepsilon_{yy}}{2}\right)(p+q) + q\varepsilon_{zz} - r\left(\frac{\varepsilon_{xx} - \varepsilon_{yy}}{2}\right)$$

$$\lambda_3 = p\varepsilon_{z'z'} + q(\varepsilon_{x'x'} + \varepsilon_{y'y'}) = p\varepsilon_{zz} + q(\varepsilon_{xx} + \varepsilon_{yy})$$

The strain-modified Raman polarizability tensors, $R_i^*$, are then determined with the eigenvectors $v_i$ of the matrix $K$ using the following relation:[69]

$$R_i^* = \sum_{k=1}^{3} v_i^{(k)} R_k$$

where $v_i^{(k)}$ denotes the $k$th component of the $i$th eigenvector. In our geometry the eigenvectors $v_i$ are given by the following expressions:

$$v_1 = \frac{1}{\sqrt{2}}\begin{bmatrix} 1 \\ -1 \\ 0 \end{bmatrix} \qquad v_2 = \frac{1}{\sqrt{2}}\begin{bmatrix} 1 \\ 1 \\ 0 \end{bmatrix} \qquad v_3 = \frac{1}{\sqrt{2}}\begin{bmatrix} 0 \\ 0 \\ 1 \end{bmatrix}$$

and thus the modified Raman tensors are:

$$R_1^* = \frac{1}{\sqrt{2}}\begin{bmatrix} 0 & 0 & -1 \\ 0 & 0 & 1 \\ -1 & 1 & 0 \end{bmatrix} \qquad R_2^* = \frac{1}{\sqrt{2}}\begin{bmatrix} 0 & 0 & 1 \\ 0 & 0 & 1 \\ 1 & 1 & 0 \end{bmatrix} \qquad R_3^* = \frac{1}{\sqrt{2}} R_3 = \frac{1}{\sqrt{2}}\begin{bmatrix} 0 & 1 & 0 \\ 1 & 0 & 0 \\ 0 & 0 & 0 \end{bmatrix}$$

**C.2 Raman selection rules in Tip Enhanced Raman Spectroscopy**

The strain induced variation of the frequency of the $q = 0$ optical phonon can be obtained by the split mode frequencies of those modes that are actually excited within the given experimental geometry as determined by the Raman selection rules.[68] In case of standard confocal far-field Raman spectroscopy in backscattered configuration only the longitudinal mode is excited.[38] In case of a near-field TERS experiment, where a metal tip in tunneling contact with the sample surface is used as an optical antenna, the Raman polarizability tensors are modified by the presence of the tip leading to a modification of the selection rules. To address the problem we used two different approaches: the first is based on a simple phenomenological model, while the second follows the electromagnetic theory of near-field Raman enhancement in a two-dimensional geometry. It is worth noting that both approaches lead to the same qualitative and quantitative results, demonstrating that in a TERS experiment in oblique backscattered configuration the signal essentially depends by the two transversal modes, which are preferentially excited (with the same probability) with respect to the longitudinal one.

*C.2.1 Simple model*

Ossikovski *et al.*[39] have recently proposed a simple model based on the introduction of a "tip-amplication tensor", $A$, accounting for the interaction of the tip with the electromagnetic field and considering that the electric field component parallel to the tip axis is preferentially amplified compared with that perpendicular to it. Assuming the tip axis along the z-axis, the tensor **A** has the following diagonal representation:

$$A = \begin{bmatrix} b & 0 & 0 \\ 0 & b & 0 \\ 0 & 0 & a \end{bmatrix}$$

where $a$ and $b$ are phenomenological tip-amplification factors (with $a > b$). They are physically related to the longitudinal and transversal tip's polarizability, respectively, and quantitatively depend on the tip geometry and material dielectric constant. The "effective" Raman polarizability tensors in presence of the tip, $R_i^{**}$, are thus obtained by the following relation:[39]

$$R_i^{**} = A^T R_i^* A,$$

which represents the action of the tip on both the incident and the scattered fields. $R_i^{**}$ are thus given by the following matrix expressions:

$$R_1^{**} = \frac{1}{\sqrt{2}}\begin{bmatrix} 0 & 0 & -ab \\ 0 & 0 & ab \\ -ab & ab & 0 \end{bmatrix} \quad R_2^{**} = \frac{1}{\sqrt{2}}\begin{bmatrix} 0 & 0 & ab \\ 0 & 0 & ab \\ ab & ab & 0 \end{bmatrix} \quad R_3^{**} = \frac{1}{\sqrt{2}}\begin{bmatrix} 0 & b^2 & 0 \\ b^2 & 0 & 0 \\ 0 & 0 & 0 \end{bmatrix}$$

The intensity $I_i$ of the scattered radiation from the $i$th mode is given by the selection rule expression:[68]

$$I_i \propto |R_i^{**} E_0|^2$$

where

$$E_0 = E_0 \begin{bmatrix} 0 \\ \cos\vartheta \\ \sin\vartheta \end{bmatrix}$$

is the electric field of the incident electromagnetic radiation (far field) and $\vartheta$ is the incidence angle of the light with respect to the z-axis. The scattered intensities for the three phonon modes are thus given by the following expressions:

$$I_1 = I_2 \propto (ab)^2 (2\sin^2\vartheta + \cos^2\vartheta)$$

$$I_3 \propto b^4 \cos^2\vartheta$$

Considering $\vartheta = 65°$ and a typical value of $a/b \approx 5.5$,[30] the ratio $I_3/I_1$ is $\approx 10^{-3}$ and thus the signal essentially depends by the two transversal modes, which are preferentially excited (with the same probability) with respect to the longitudinal one.

*C.2.2 Electromagnetic theory of field enhancement*

In this section the Raman selection rules are derived by developing an electromagnetic theory of the field enhancement following the formalism reported by Cancado *et al.*.[70] Fig. 10(a) shows the experimental configuration and the coordinates used in the theoretical analysis. The electric field of the incident electromagnetic radiation, $\boldsymbol{E_0}$, is amplified and localized by the tip. The total electric field is thus represented by the superposition of the incident laser field and of the localized field generated by the tip acting as an optical antenna. Nearby the tip the electric field resembles the field of an induced dipole located at the center of a small sphere of radius $R$:

$$\boldsymbol{E}(\hat{\boldsymbol{r}}, \omega) \approx \boldsymbol{E_0}(\boldsymbol{r}, \omega) + \frac{\omega^2}{\varepsilon_0 c^2} \boldsymbol{G^0}(\boldsymbol{r}, \hat{\boldsymbol{r}}; \omega) \boldsymbol{\alpha}_{tip}(\omega) \boldsymbol{E_0}(\boldsymbol{r}, \omega)$$

where $\boldsymbol{G^0}$ represents the free-space dyadic Green's function in absence of a tip, $\boldsymbol{\alpha}_{tip}$ is the tip's polarizability, and $c$ is the vacuum speed of the light. In the following we can neglect the field $\boldsymbol{E_0}$ in the previous equation since the near field due to the tip is generally much stronger. The tip's polarizability has the following matrix representation:

$$\boldsymbol{\alpha}_{tip} = \frac{4}{3}\pi R^3 \begin{bmatrix} \frac{\varepsilon - 1}{\varepsilon + 2} & 0 & 0 \\ 0 & \frac{\varepsilon - 1}{\varepsilon + 2} & 0 \\ 0 & 0 & f_e \end{bmatrix}$$

in the *xyz* stripe reference framework, where $f_e$ is the field-enhancement factor and the transverse polarizability components correspond to the quasi-static polarizability of a small sphere. The free-space dyadic Green's function is defined in the electromagnetic theory by the electric field generating by a radiating electric dipole.[71] The near-field term can be written as:

$$\boldsymbol{G^0}(\boldsymbol{r}, \hat{\boldsymbol{r}}; \omega) = \frac{e^{ik\rho}}{4\pi k^2 \rho^3} \left[ -\mathbf{I} + \frac{3\boldsymbol{\rho}\boldsymbol{\rho}}{\rho^2} \right]$$

where $k = \omega/c$ is the wavevector of the electromagnetic field in the free-space, $\mathbf{I}$ is the unit dyad, $\rho$ is the absolute value of the vector $\boldsymbol{\rho} = \boldsymbol{r} - \hat{\boldsymbol{r}}$ and $\boldsymbol{\rho}\boldsymbol{\rho}$ is the dyadic product of $\boldsymbol{\rho}$ with itself. Considering the coordinate system defined in Fig. 10(a) and the stripe geometry, the relative position vector $\boldsymbol{\rho}$ can be written in Cartesian coordinate as:

$$\boldsymbol{\rho} = x\boldsymbol{x} + y\boldsymbol{y} + (\Delta + R)\boldsymbol{z},$$

and thus the matrix components of the dyadic Green's function can be evaluated:

$$G^0_{xx} = \frac{e^{ik\rho}}{4\pi k^2 \rho^5}[2x^2 - y^2 - (\Delta + R)^2]$$

$$G^0_{yy} = \frac{e^{ik\rho}}{4\pi k^2 \rho^5}[2y^2 - x^2 - (\Delta + R)^2]$$

$$G_{zz}^0 = \frac{e^{ik\rho}}{4\pi k^2 \rho^5}[2(\Delta + R)^2 - x^2 - y^2]$$

$$G_{xz}^0 = G_{zx}^0 = \frac{e^{ik\rho}}{4\pi k^2 \rho^5} 3x(\Delta + R)$$

$$G_{yz}^0 = G_{zy}^0 = \frac{e^{ik\rho}}{4\pi k^2 \rho^5} 3y(\Delta + R)$$

$$G_{xy}^0 = G_{yx}^0 = \frac{e^{ik\rho}}{4\pi k^2 \rho^5} 3xy$$

Panels (b)-(g) in Fig. 10 show a graphic representation of the six independent tensor components $G_{xx}^0$, $G_{yy}^0$, $G_{zz}^0$, $G_{xy}^0$, $G_{xz}^0$, and $G_{yz}^0$ in the *xy* plane using $\Delta = 0.7$ nm and $R = 20$ nm.

The localized field generated by the laser-irradiated metal tip interacts with the nano-stripe inducing an electric dipole $\boldsymbol{p}(\hat{\boldsymbol{r}}, \omega_s)$:

$$\boldsymbol{p}(\hat{\boldsymbol{r}}, \omega_s) \propto \boldsymbol{R}_i^*(\hat{\boldsymbol{r}}, \omega, \omega_s) \boldsymbol{E}(\boldsymbol{r} - \hat{\boldsymbol{r}}, \omega)$$

where $\omega_s$ is the shifted Raman frequency and $\boldsymbol{R}_i^*$ are the strain-modified Raman polarizability tensors of the three modes (where $i = 1$ (TO1), 2 (TO2), and 3 (LO)). The scattered field at the detector with shifted frequency $\omega_s$ is the field generated by the tip's secondary sources induced by the Raman dipole distribution $\boldsymbol{p}(\hat{\boldsymbol{r}}, \omega_s)$. In approximation of leading dipole term, the scattered field is thus obtained as:

$$\boldsymbol{E}_i^s(\boldsymbol{r}, \omega_s) \propto \int_{-\infty}^{+\infty} d\hat{x} \int_{-\infty}^{+\infty} d\hat{y}\, \boldsymbol{\alpha}_{tip}(\omega_s) \boldsymbol{G}^0(\boldsymbol{r}, \hat{x}, \hat{y}; \omega_s) \boldsymbol{p}(\hat{x}, \hat{y}, \omega_s)$$

$$\approx \int_{-\infty}^{+\infty} d\hat{x} \int_{-\infty}^{+\infty} d\hat{y}\, \boldsymbol{\alpha}_{tip}(\omega_s) \boldsymbol{G}^0(\boldsymbol{r}, \hat{x}, \hat{y}; \omega_s) \boldsymbol{R}_i^*(\hat{x}, \hat{y}, \omega, \omega_s) \boldsymbol{G}^0(\boldsymbol{r}, \hat{x}, \hat{y}; \omega) \boldsymbol{\alpha}_{tip}(\omega) \boldsymbol{E}_0(\boldsymbol{r}, \omega)$$

Using the given tensor representations for the tip's polarizability, the free-space dyadic Green's function and the strain-modified Raman polarizability, we obtain:

$$\boldsymbol{E}_1^s \propto \int_{-\infty}^{+\infty} d\hat{x} \int_{-\infty}^{+\infty} d\hat{y} \begin{bmatrix} -f_e(\omega) \frac{\varepsilon(\omega_s) - 1}{\varepsilon(\omega_s) + 2} \sin\vartheta\, (G_{xz}^0 G_{xz}^0 + G_{xx}^0 G_{zz}^0) \\ f_e(\omega) \frac{\varepsilon(\omega_s) - 1}{\varepsilon(\omega_s) + 2} \sin\vartheta\, (G_{yz}^0 G_{yz}^0 + G_{yy}^0 G_{zz}^0) \\ f_e(\omega_s) \frac{\varepsilon(\omega) - 1}{\varepsilon(\omega) + 2} \cos\vartheta\, (G_{yz}^0 G_{yz}^0 + G_{yy}^0 G_{zz}^0) \end{bmatrix}$$

$$E_2^s \propto \int_{-\infty}^{+\infty} d\hat{x} \int_{-\infty}^{+\infty} d\hat{y} \begin{bmatrix} f_e(\omega) \dfrac{\varepsilon(\omega_s) - 1}{\varepsilon(\omega_s) + 2} \sin\vartheta \, (G_{xz}^0 G_{xz}^0 + G_{xx}^0 G_{zz}^0) \\ f_e(\omega) \dfrac{\varepsilon(\omega_s) - 1}{\varepsilon(\omega_s) + 2} \sin\vartheta \, (G_{xz}^0 G_{xz}^0 + G_{xx}^0 G_{zz}^0) \\ f_e(\omega_s) \dfrac{\varepsilon(\omega) - 1}{\varepsilon(\omega) + 2} \cos\vartheta \, (G_{yz}^0 G_{yz}^0 + G_{yy}^0 G_{zz}^0) \end{bmatrix}$$

$$E_3^s \propto \int_{-\infty}^{+\infty} d\hat{x} \int_{-\infty}^{+\infty} d\hat{y} \begin{bmatrix} \dfrac{\varepsilon(\omega) - 1}{\varepsilon(\omega) + 2} \dfrac{\varepsilon(\omega_s) - 1}{\varepsilon(\omega_s) + 2} \cos\vartheta \, (G_{xy}^0 G_{xy}^0 + G_{xx}^0 G_{xx}^0) \\ 0 \\ 0 \end{bmatrix}$$

The intensity $I_i$ of the scattered radiation from the $i$th mode is given by:

$$I_i \propto |E_i^s(r, \omega_s)|^2,$$

and thus:

$$I_1 = I_2 \propto g_{1,2} \left\{ 2\left[ f_e(\omega) \frac{\varepsilon(\omega_s) - 1}{\varepsilon(\omega_s) + 2}\right]^2 \sin^2\vartheta + \left[ f_e(\omega_s) \frac{\varepsilon(\omega) - 1}{\varepsilon(\omega) + 2}\right]^2 \cos^2\vartheta \right\}$$

$$I_3 \propto g_3 \left[\frac{\varepsilon(\omega) - 1}{\varepsilon(\omega) + 2}\right]^2 \left[\frac{\varepsilon(\omega_s) - 1}{\varepsilon(\omega_s) + 2}\right]^2 \cos^2\vartheta$$

where

$$g_{1,2} = \int_{-\infty}^{+\infty} d\hat{x} \int_{-\infty}^{+\infty} d\hat{y} \, (G_{yz}^0 G_{yz}^0 + G_{yy}^0 G_{zz}^0) = \int_{-\infty}^{+\infty} d\hat{x} \int_{-\infty}^{+\infty} d\hat{y} \, (G_{xz}^0 G_{xz}^0 + G_{xx}^0 G_{zz}^0)$$

$$g_3 = \int_{-\infty}^{+\infty} d\hat{x} \int_{-\infty}^{+\infty} d\hat{y} \, (G_{xy}^0 G_{xy}^0 + G_{xx}^0 G_{xx}^0)$$

with $g_3 \approx 0.41 g_{1,2}$ considering $\Delta = 0.7$ nm and $R = 20$ nm.

In case of tip enhanced scattering the polarizability of the metal tip is much stronger along its axis than in the transversal plane ($f_e > (\varepsilon - 1)/(\varepsilon + 2)$), then the intensity of the longitudinal mode, $I_3$, is much lower than that of the two transversal modes, $I_1$ and $I_2$. Moreover, in hypothesis that $f_e(\omega) \approx f_e(\omega_s)$ and $\varepsilon(\omega) \approx \varepsilon(\omega_s)$ since $\omega \approx \omega_s$, we obtain the same quantitative results as found from the simple phenomenological model considering that $a \propto f_e(\omega)$ and $b \propto \dfrac{\varepsilon(\omega)-1}{\varepsilon(\omega)+2}$.

*C.2.3 Quantum mechanical approach*

The results obtained in the previous sections are also fully consistent with a quantum mechanical expression of the Raman selection rules. Breaking of the translational symmetry due to the presence of the surface lowers the symmetry of the crystal from spherical to cylindrical, and thus only the projection of the angular momentum along the z-axis, $L_z$, is conserved during the photon-phonon scattering process:[72]

$$m'_{photon} + m''_{photon} - m_{phonon} = 0$$

where $m'_{photon}$, $m''_{photon}$, and $m_{phonon}$ are the $L_z$ quantum numbers for the incident photon, the scattered photon, and the phonon, respectively. Since the metal tip, which represents the nearfield source, can be assimilated to an emitting electric dipole, both incident and scattered photons have null $L_z$ quantum numbers: $m'_{photon} = m''_{photon} = 0$, and thus $m_{phonon} = 0$. This means that a strong coupling between the near-field radiation and the crystal lattice is allowed only for lattice vibrations with polarization along the z-axis and angular and linear momenta completely lying in the *xy* plane, i.e. only for transversal modes.

**C.3 Strain determination**

The frequency splitting for the *i*th mode, $\omega_i$, is related to the corresponding eigenvalue $\lambda_i$ of the **K** matrix by the following expression:

$$\omega_i^2 = \omega_0^2 + \lambda_i$$

and thus

$$\Delta\omega_i = (\omega_i - \omega_0) \approx \frac{\lambda_i}{2\omega_0}$$

where $\omega_0$ is the unstrained mode frequency. From the Raman selection rules in TERS case we derived that the $q = 0$ Raman optical phonon peaks in the measured TERS spectra are featured by a superposition of the only two lineshapes associated to the transversal modes TO1 and TO2. Since they are excited with the same probability, the strain induced frequency shift, $\Delta\omega = \omega - \omega_0$, can be simply obtained by the average of the TO1 and TO2 mode frequency splittings:[73,74]

$$\Delta\omega \approx \frac{\Delta\omega_1 + \Delta\omega_2}{2} \approx \frac{\lambda_1 + \lambda_2}{4\omega_0}$$

and then, using the strain dependent expression of $\lambda_1$ and $\lambda_2$ (see Section S.2.1):

$$\Delta\omega \approx \frac{1}{2\omega_0}\left[\left(\frac{\varepsilon_{xx} + \varepsilon_{yy}}{2}\right)(p + q) + q\varepsilon_{zz}\right]$$

This relation can be conveniently expressed as a function of the trace of the strain tensor $(\varepsilon_{xx} + \varepsilon_{yy} + \varepsilon_{zz})$, thus:

$$\Delta\omega \approx \frac{1}{2\omega_0}\left[\left(\frac{p+q}{2}\right)(\varepsilon_{xx}+\varepsilon_{yy}+\varepsilon_{zz})+\left(\frac{q-p}{2}\right)\varepsilon_{zz}\right]$$

The phonon deformation potentials $p$ and $q$ can be then represented using adimensional quantity ($p = K_{11}\omega_0^2$ and $q = K_{12}\omega_0^2$):

$$\Delta\omega \approx \frac{\omega_0}{2}\left[\left(\frac{K_{11}+K_{12}}{2}\right)(\varepsilon_{xx}+\varepsilon_{yy}+\varepsilon_{zz})+\left(\frac{K_{12}-K_{11}}{2}\right)\varepsilon_{zz}\right]$$

Considering that $K_{11} \approx K_{12}$ for high Ge content SiGe alloy[38,41] (as in the case of the nano-stripes studied in this work), we can neglect the second term on the right side of the previous equation, leading to the following expression of the strain induced frequency shift:

$$\Delta\omega \approx \frac{\omega_0}{2}\left[\left(\frac{K_{11}+K_{12}}{2}\right)(\varepsilon_{xx}+\varepsilon_{yy}+\varepsilon_{zz})\right]$$

which can be more conveniently expressed as a function of the hydrostatic strain $\varepsilon_h = \frac{1}{3}(\varepsilon_{xx}+\varepsilon_{yy}+\varepsilon_{zz})$:

$$\Delta\omega \approx \frac{3\omega_0}{4}(K_{11}+K_{12})\varepsilon_h$$

finally obtaining the Eq. (2) reported in the main text.

## APPENDIX D: DFT-LDA CALCULATIONS

We addressed the structural and electronic properties of Ge(001) by means of the density-functional theory (DFT), using the plane-wave pseudopotential method and the local density approximation (LDA) for the exchange-correlation potential, whose accuracy have been demonstrated in a variety of systems.[75,76] The well-known tendency of DFT-LDA to underestimate the gaps of semiconductors[77] is not a significant issue for the present calculations, where we address mainly *changes* in band energies as a function of (small) strains. Disagreement with experimental determinations of the strain variation of the valence bandwidth were reported,[78] but are probably less relevant for the positions of levels near the Fermi energy. We adopt a Perdew-Zunger LDA functional,[79] a norm-conserving scalar-relativistic pseudopotential, and a 30 Ry (408 eV) cutoff for the plane-wave basis.

The simulations are carried out in a standard supercell geometry. The conventional cell bulk lattice parameter is $a = 561.6$ pm, as obtained by a full relaxation of the bulk structure, slightly shorter than experiment. The supercell is 2×1 to accommodate the b(2×1) surface reconstruction.[80] The *k*-point mesh involves 4×8 points, including Γ. The slab representing the surface consists of 10 fixed layers plus 3 surface layers on each side, where atoms are fully allowed to relax in all directions, until all force components are smaller than 0.4 pN. All computed band energies are referred to the vacuum reference potential outside the solid. This reference potential is determined by averaging the Hartree potential over a vacuum region at the middle between two copies of the periodically repeated 2.15 nm-thick Ge slab. A relatively thick (2.34 nm) vacuum region between periodic copies of the Ge slab makes the potential almost costant (within 0.01 meV) over a 1 nm -thick region there. Strain is applied by increasing the cell sides by 1%. The atomic positions of the surface layers relax in all strained geometries, which implies that the z-oriented strain affects mostly the bulk layers. While the calculations are carried out for intrinsic Ge, as the Ge stripes are immersed in a n+ Si host, the Fermi level is to be taken to be pinned at the lowest bulk conduction states. Accordingly, the most reliable estimate of the work function at all regimes of strain is the one obtained by the position of these bulk band states referred to the vacuum level. For a tensile strained Ge slab, we found a reduction of the surface electrostatic dipole at the surface, but at the same time a shift of both the Fermi level and the conduction band opposite to the surface dipole decrease.

It is worth noting that DFT calculations have been performed considering a perfectly ordered surface. However, on a real surface local disorder and defects allow relaxation of the atomic density toward its ideal value. Thus, the surface dipole could be less sensitive to the strain than predicted by the calculations, leading to a higher work function shift. Moreover an additional surface barrier, not directly taken into account in the DFT calculations, is experienced by an electron escaping from the nano-stripe surface. This barrier is induced by a negative charge layer formed at the nano-stripe surface due to a redistribution of the space charge accumulating at the Si/SiGe interface as a consequence of the band bending following the thermal equilibrium condition at the heterojunction. Considering the given doping level and Si/Ge natural band offset we estimate that the order of magnitude of this barrier is around 10 meV, which is within the experimental uncertainty for the measured work function value. This suggests that both strain and space charge arguments are consistent with an increase of the work function for the nano-stripes with respect to the bulk case, as experimentally obtained by XPEEM, and that our DFT calculations are thus able to correctly describe the strain-induced electronic structure changes inside the stripe.

**APPENDIX E: SENSITIVITY ANALYSIS OF FEM SIMULATIONS**

In this section we address whether the idealized shape of the interface between the SiGe stripe and the Si substrate, used in the theoretical FEM simulations and exploited to extract the strain tensor, could significantly alter the results. This is particularly important in the present scenario where the relevance of our experimental work relies on its consistency with the Finite Element Modeling (FEM). Thus we have devoted some extra-simulations to further elucidate this point.

The hydrostatic strain map reported in Fig. 5(b) is represented again in Fig. 11(a), on a strain scale convenient to perform the comparison that follows. Let us start from a simple observation: if we remove the whole dislocation net, we obtain the new map shown in Fig. 11(b). Obviously, the strain field close to the lower interface is very different as elastic relaxation only has little effect, so that in the absence of linear defects the misfit strain is basically maintained. However, the hydrostatic strain close to the upper free surface, i.e. the only one relevant for nanoelectronics tensor application and detected by TERS, is tensile exactly as in Fig. 11(a), and also rather close in value. We recall that one of the main points of the paper is to show that the present nano-structuring allows for a hydrostatic tensile strain significantly higher than the typical thermal one in flat films (of the order of 0.1%). In this respect, to get the main effect one does not need to explicitly consider dislocations. This allows us to investigate systematically the effect of the stripe shape and interface profile, as elastic calculations are much faster than the combined elastic-plastic simulations as the one reported in Fig. 11(a). So, in Figs. 11(c)-(f) we considered different shapes and profiles: the structures shown in (c) and (d) are perhaps closer to the experimental one reported in Fig. 1, in (e) we are considering a fully rounded shape, while in (f) we report the results for the case of a wavy interface. In all shown simulations, the hydrostatic strain close to the upper free surface is clearly always of the same order, around 0.3%. Accordingly, we deduce that the main conclusions of our work depend very mildly on:

a)   the detailed dislocation distribution/density, etc.

b)   the actual details of the stripe shape and interface profile (obviously, for a given vertical-to-horizontal aspect ratio).

Thus our choice of an idealized rectangular shape seems to be well justified.

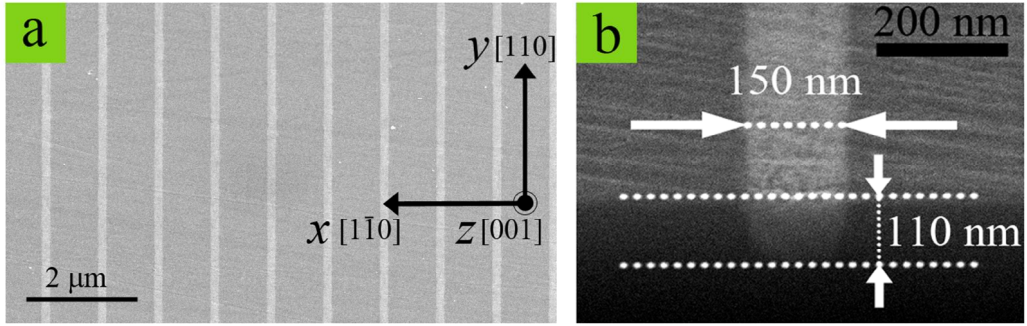

FIG. 1. Laterally confined SiGe nano-stripes. Panel (a) Top view SEM image of the periodic array of nano-stripes. Panel (b) Cross-section SEM image of a single nano-stripe after focused ion beam (FIB) processing. During FIB processing the ion beam hits the sample surface with an incidence angle of 52° with respect to the normal direction. In this condition the cross-section profile of the nano-stripe can be well distinguished by possible ion-induced artifacts due to the amorphization of the cross-section surface, which could appear only along the 52° tilted direction.

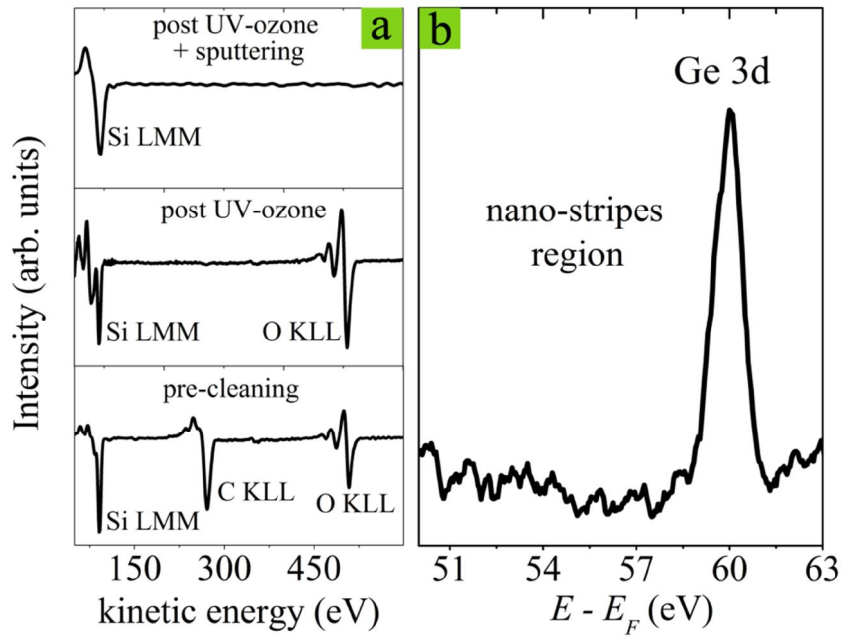

FIG. 2. Sample cleaning for XPEEM experiment. (a): Auger spectra monitoring the surface contamination during the different steps of the cleaning procedure on a Si test sample: pre-cleaning (bottom), post UV-ozone treatment (centre), and post sputtering (top). (b): photoemission spectrum of the Ge 3d core level measured on the nano-stripes after the cleaning procedure. The absence of any shifted structure appears at the low kinetic energy side indicates the absence of germanium-oxide.

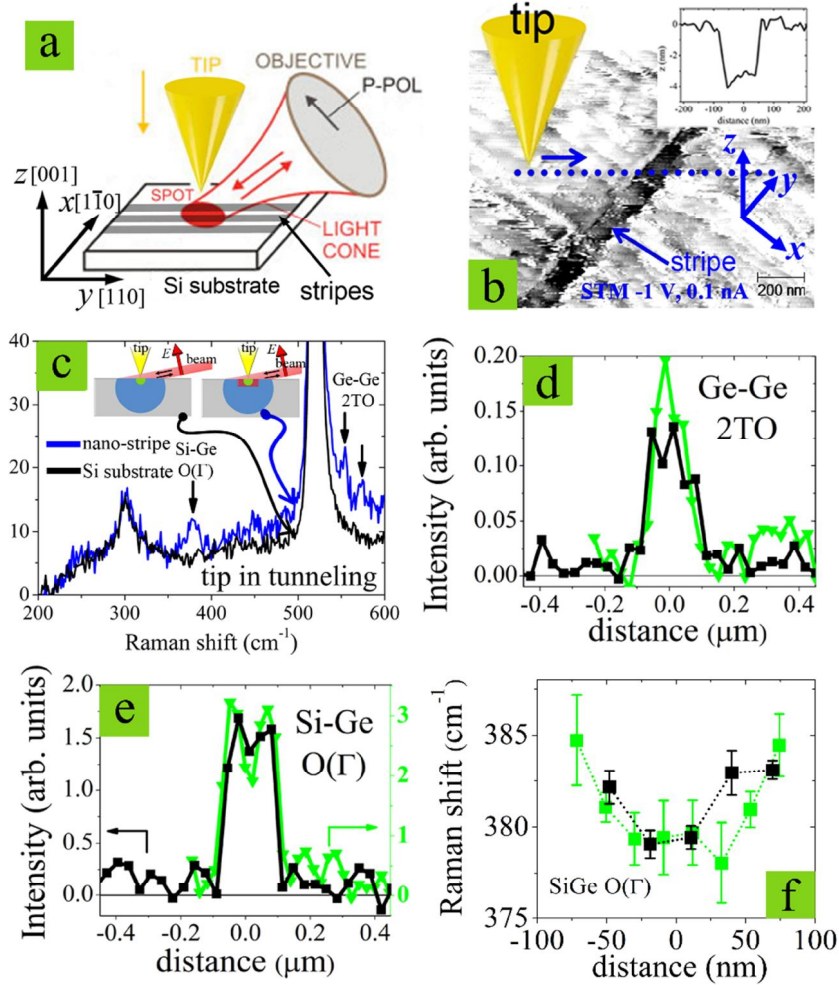

FIG. 3. TERS data. Panel (a): schematic representation of the TERS experiment; incident light has a *p*-polarization and the scattering plane is represented by the *yz* plane. Panel (b): STM image of a single SiGe nano-stripe. The dotted blue line represents the path over which the tip is scanned and Raman spectra were acquired. Inset: STM cross-section profile across the nano-stripe. Panel (c): baseline corrected TERS spectra measured with the STM tip in tunneling on the SiGe nano-stripe (blue line) and on the Si substrate (black line). Insets: scheme of the experimental geometry for the spectra shown in the main panels; the far-field and the near-field interaction volumes within the sample are represented by the blue and green areas, respectively; the electric field

polarization (*E*) and the scattering directions are defined by the red and black arrows, respectively. Panels (d) and (e): background subtracted intensity profiles of the Ge-Ge 2TO and the Si-Ge first order peaks, respectively, as derived by TERS spectra monitored as a function of the position across a single nano-stripe along two scan lines of the tip acquired at two different position along the stripe axis. Panel (f): Raman frequency profiles of the SiGe mode as a function of the position across the stripe for the two scan lines shown in the panels (d) and (e).

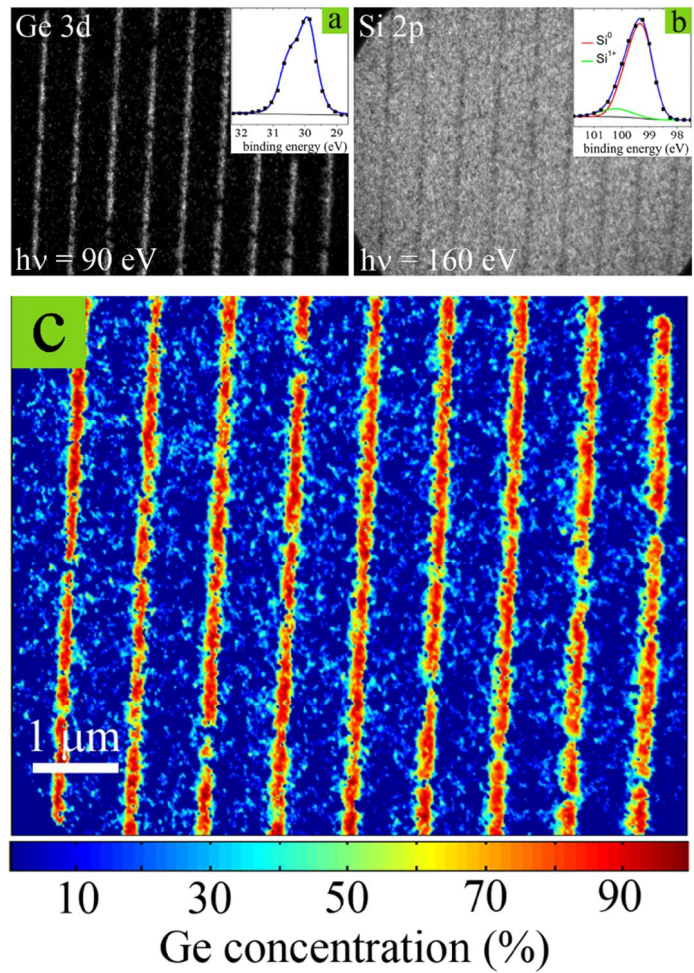

FIG. 4. Ge concentration mapping. Panel (a)-(b): background subtracted Ge 3d and Si 2p core level XPEEM images (FoV is ~ 12 μm) of the nano-stripes array; insets: Ge 3d and Si 2p photoemission spectra extracted on a single nano-stripe (black squares) fitted with Voigt lineshape components (solid lines). In the case of Ge 3d spectrum two spin-orbit split structures separated by 0.6 ± 0.1 eV and with a branching ratio of ~ 1.5 have been considered. The weak component at high binding energy side within the Si 2p spectrum is consistent with a 1+

ionization state. Panel (c): spatial map of the Ge concentration as obtained by monitoring the pixel-by-pixel Ge 3d and Si 2p peak peaks and fitting their intensities with a standard quantification model.

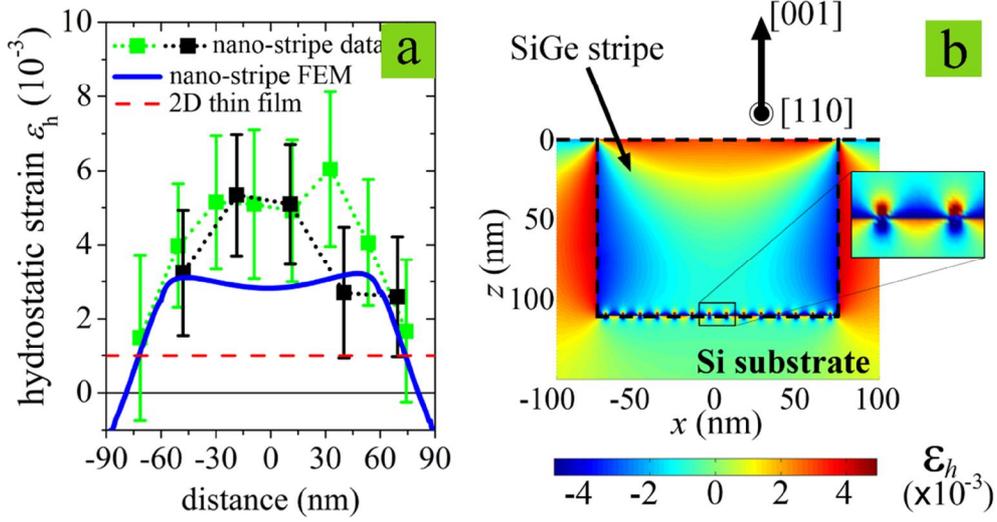

FIG.5 Panel (a): experimental spatial profiles (black and green squares) of the hydrostatic strain, $\varepsilon_h$, obtained combining TERS and XPEEM data. The blue solid line represents the computed $\varepsilon_h$ profile as obtained by FEM simulations (see text). The red dashed line is the hydrostatic strain value in the case of a 2D thin film. Panel (b): spatial map of the hydrostatic deformation in the $xz$ plane obtained by FEM simulations. Inset: strain field created by two 90° dislocations at the interface between the SiGe stripe and the Si substrate.

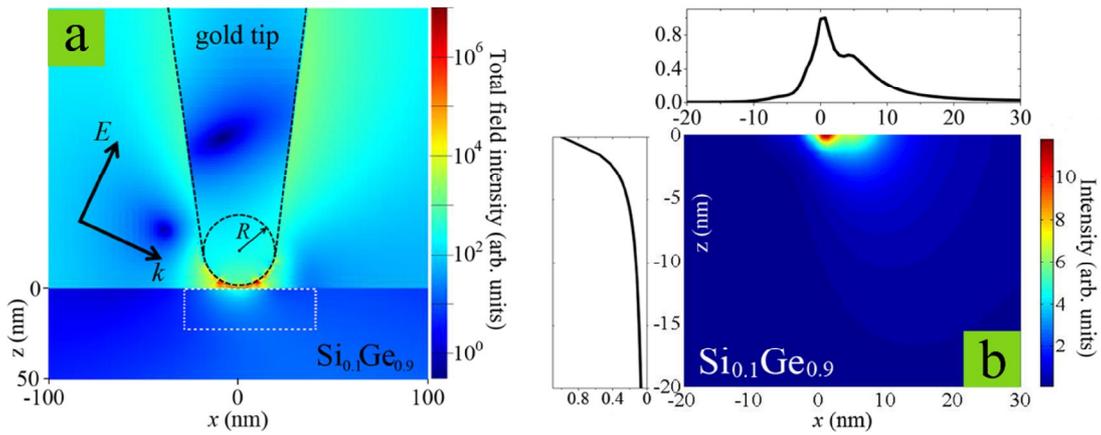

FIG. 6. FDTD simulation of TERS experiment. Panel (a): intensity of total electric field in the $xz$ plane after interaction with a gold tip having a radius of 20 nm and separated of 0.7 nm from the $z = 0$ plane (the (001)

plane), calculated with a finite-difference time-domain (FDTD) solver. Panel (b): spatial map of the Raman signal (proportional to the fourth power of the electric field) in a $Si_{0.1}Ge_{0.9}$ alloy in the region defined by the dotted white rectangle shown in panel (a). The electromagnetic radiation from the laser is coming from the left side. Insets: transversal (top) and longitudinal (left) profiles of the Raman intensity.

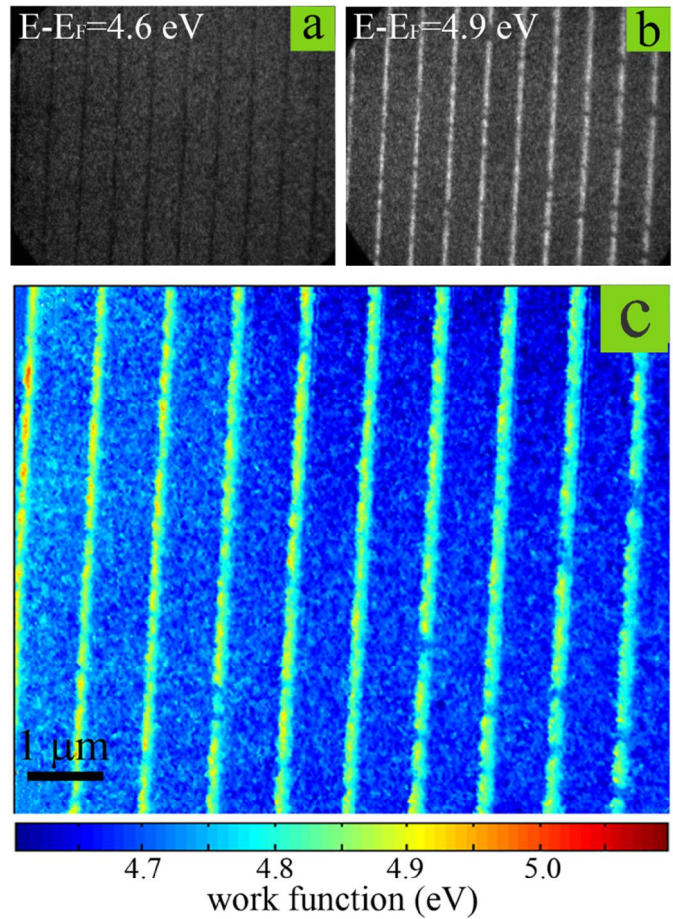

FIG. 7. Work function mapping. Panels (a)-(b): XPEEM images of the nano-stripes array acquired with soft x-ray excitation at hν = 90 eV using secondary electrons of 4.6 eV (a) and 4.9 eV (b). The FoV is ~15 µm. Panel (c): Local work function map obtained from the experimental photoemission threshold spectra taken pixel by pixel and least-square-fitted to the secondary electron distribution described by Henke's model (see Appendix B).

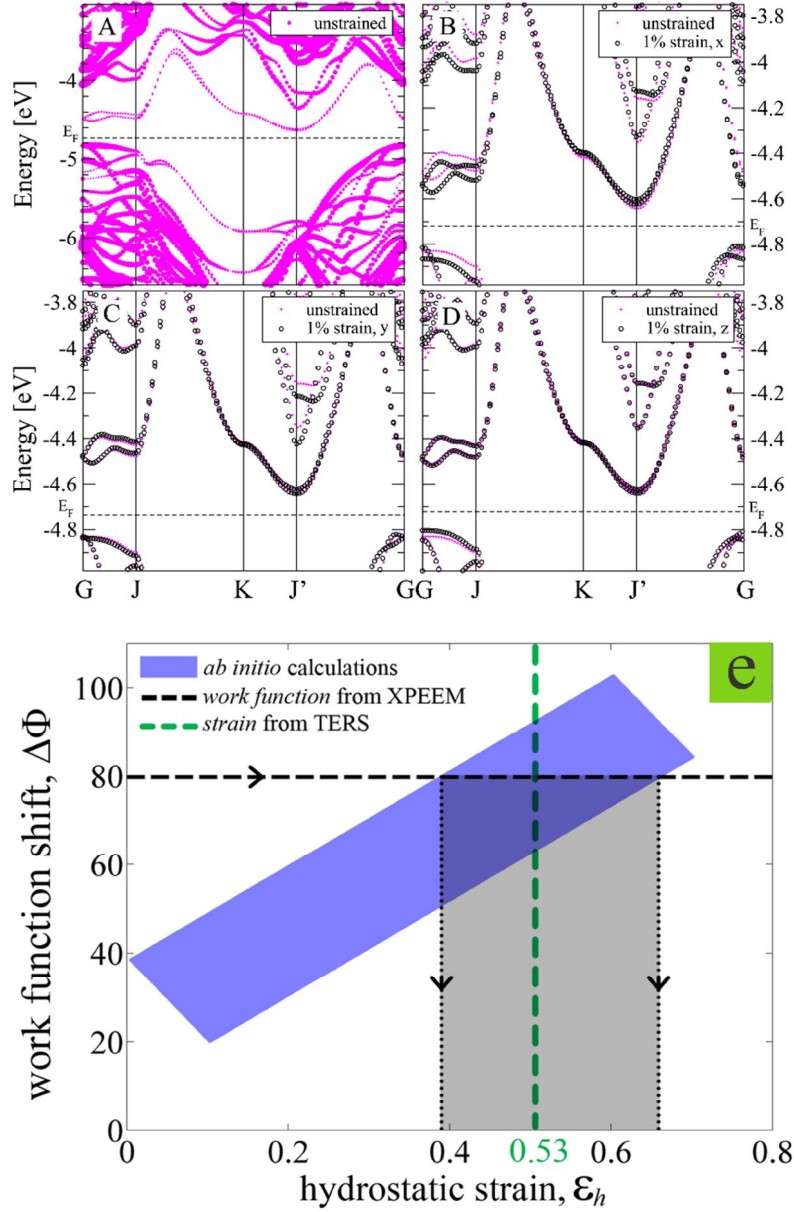

FIG. 8. DFT-LDA: work function calculation. Calculated band structure of the Ge(001)b(2×1) surface using density functional theory in LDA approximation under different tensile strain conditions. In all panels the magenta crosses (×) represent the unstrained band structure, while the black circles (○) describe the effect of the applied tensile strain. The dashed line defines the Fermi level. Panel (a): Unstrained surface, the size of points is proportional to the bulk partial weight of those bands states:

thin states are surface bands. Panel (b): $\varepsilon_{xx} = 0.01$, $\varepsilon_{yy} = 0$ and $\varepsilon_{zz} = 0$. Panel (c): $\varepsilon_{xx} = 0$, $\varepsilon_{yy} = 0.01$ and $\varepsilon_{zz} = 0$. Panel (d) $\varepsilon_{xx} = 0$, $\varepsilon_{yy} = 0$ and $\varepsilon_{zz} = 0.01$. When in contact with n+ silicon, the Fermi energy of the Ge stripe moves up to pin the bulk states near -4 eV. Panel (e): calculated work function work function shift, $\Delta\Phi$, as a function of the hydrostatic strain, $\varepsilon_h$, inside the nano-stripes (see text for details). The comparison of the experimentally measured work function change (black dashed line) with the calculated values, gives an estimation of the strain state of the nano-stripe (grey region). The dashed green line represents the strain value as measured by TERS (~ 0.53 %).

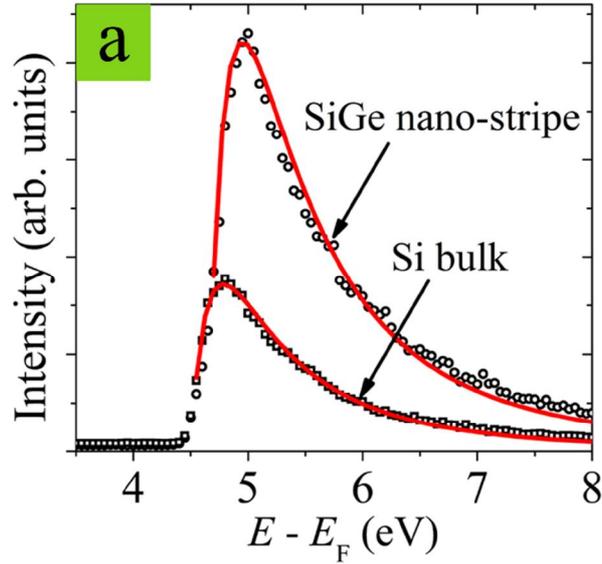

FIG. 9. Experimental secondary electron energy distributions (open squares and circles) as a function of $E - E_F$ for the Si bulk and the SiGe nano-stripes, respectively. The red curves represent the best least-square fitting of the experimental data using the Henke's model.

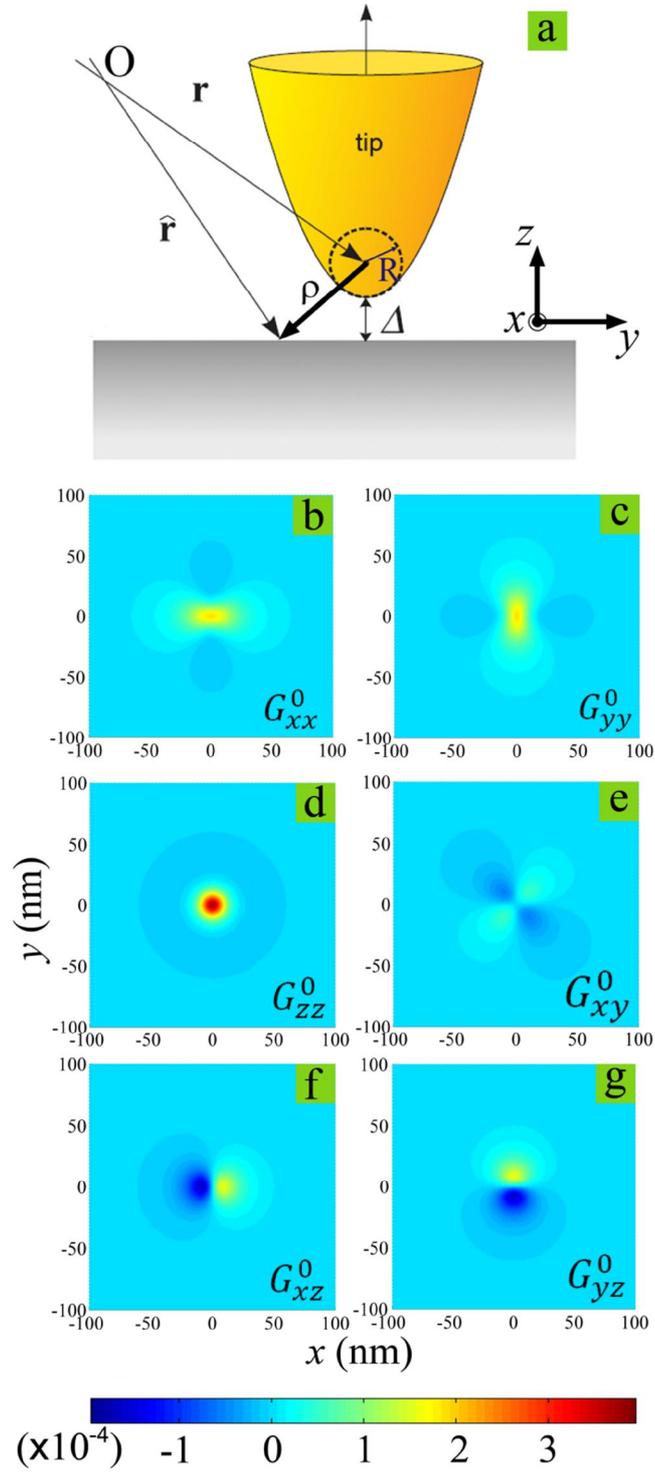

FIG. 10. Dyadic Green's functions in TERS geometry. (a): TERS experimental configuration and system of coordinates used in the theoretical analysis of field enhancement and Raman selection rules. (b)-(g): graphic

representation of the six independent tensor components $G^0_{xx}$, $G^0_{yy}$, $G^0_{zz}$, $G^0_{xy}$, $G^0_{xz}$, and $G^0_{yz}$ of the dyadic Green's function plotted in the *xy* plane considering $\Delta = 0.7$ nm and $R = 20$ nm.

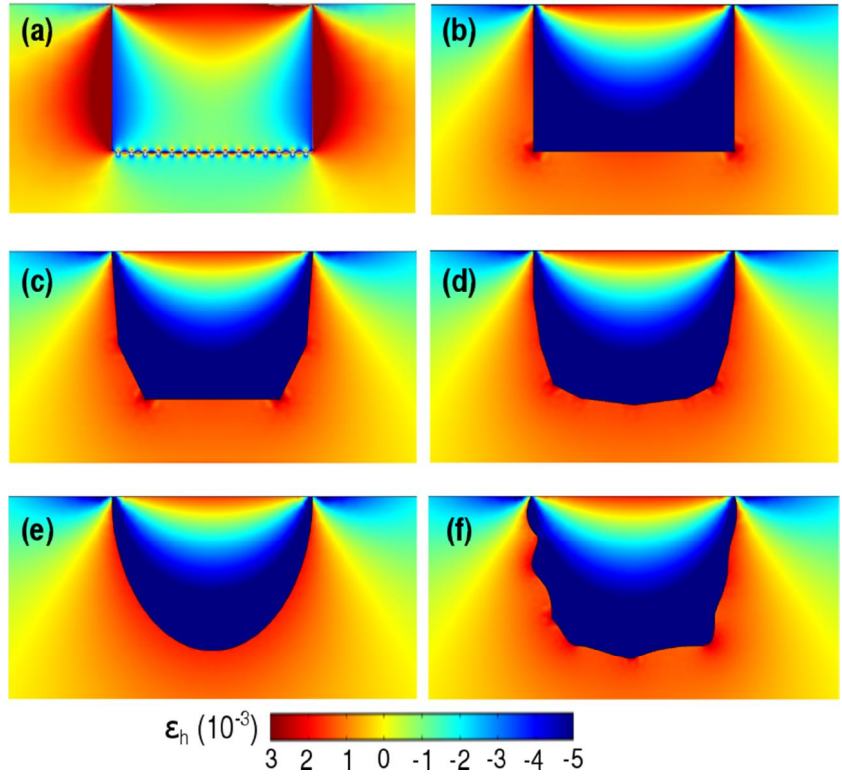

FIG. 11. Sensitivity analysis of FEM simulations. Simulated hydrostatic strain maps as obtained by Finite Element Modeling (FEM) for different relaxation conditions, stripe shapes and interface profiles: (a) both elastic and plastic (formation of dislocation) relaxation mechanisms are considered for a rectangular stripe shape; (b) rectangular shape with only elastic relaxation; (c)-(d) elastic relaxation for polygonal shape profiles (closer to the experimental shape reported in Fig. 1 of the main text); (e) elastic relaxation on a fully rounded shape; (f) elastic relaxation for the case of a wavy interface profile.